# Voltage-controlled electron tunnelling from a single self-assembled quantum dot embedded in a two-dimensional-electron-gas-based photovoltaic cell

J. D. Mar[1,2,*], X. L. Xu[1], J. J. Baumberg[2], A. C. Irvine[3], C. Stanley[4], and D. A. Williams[1]

[1]*Hitachi Cambridge Laboratory, Cavendish Laboratory, Cambridge CB3 0HE, United Kingdom*

[2]*NanoPhotonics Centre, Cavendish Laboratory, University of Cambridge, Cambridge CB3 0HE, United Kingdom*

[3]*Microelectronics Research Centre, Cavendish Laboratory, University of Cambridge, Cambridge CB3 0HE, United Kingdom*

[4]*Department of Electronics & Electrical Engineering, University of Glasgow, Glasgow G12 8QQ, United Kingdom*

**ABSTRACT**

We perform high-resolution photocurrent (PC) spectroscopy to investigate resonantly the neutral exciton ground-state ($X^0$) in a single InAs/GaAs self-assembled quantum dot (QD) embedded in the intrinsic region of an *n-i*-Schottky photodiode based on a two-dimensional electron gas (2DEG), which was formed from a Si $\delta$-doped GaAs layer. Using such a device, a single-QD PC spectrum of $X^0$ is measured by sweeping the bias-dependent $X^0$ transition energy through that of a fixed narrow-bandwidth laser via the quantum-confined Stark effect (QCSE). By repeating such a measurement for a series of laser energies, a precise relationship between the $X^0$ transition energy and bias voltage is then obtained. Taking into account power broadening of the $X^0$ absorption peak, this allows for high-resolution measurements of the $X^0$ homogeneous linewidth and, hence, the electron tunnelling rate. The electron tunnelling rate is measured as a function of the vertical electric field and described accurately by a theoretical model, yielding information about the electron confinement energy and QD height. We demonstrate that our devices can operate as 2DEG-based QD photovoltaic cells and conclude by proposing two optical spintronic devices that are now feasible.

PACS numbers: 78.67.Hc, 73.21.La, 73.23.Hk, 73.50.Pz

*Author to whom correspondence should be addressed. Electronic mail: jm585@cam.ac.uk

# I. INTRODUCTION

The three-dimensional quantum confinement of electron-hole (*e-h*) pairs into discrete energy levels in self-assembled quantum dots (QDs) has motivated research into the incorporation of these so-called "artificial atoms" into next-generation quantum optoelectronic devices for the implementation of quantum information processing. Such landmark devices include charge-tunable QD diodes,[1,2] electrically-pumped single-photon sources,[3,4] and single-QD photodiodes.[5,6] A common characteristic of all these devices is that they are able to serve as an interface between optical fields and an electronic circuit by taking advantage of the ability of self-assembled QDs to couple to light efficiently. Specifically, single-QD photodiodes have been used to measure electrically the coherent optical manipulation of a QD two-level system.[6,7] Here, the state of a two-level system is read-out electrically as the amplitude of a photocurrent signal created by exciton ionization and tunnelling of optically-excited *e-h* pairs from the QD into an electronic circuit connected across the photodiode. However, until now, all work that has measured photocurrent from a single QD has employed a photodiode structure based on an *n*-type doped *bulk*-layer.[5-7]

Here, we demonstrate voltage-controlled electron tunnelling from a single self-assembled QD embedded in the intrinsic region of an *n-i*-Schottky photodiode based on a two-dimensional electron gas (2DEG). A high-resolution photocurrent (PC) spectrum of the neutral ground-state QD exciton ($X^0$) is obtained by sweeping the $X^0$ transition energy through that of a fixed laser via the quantum-confined Stark effect (QCSE). By repeating this measurement for a series of laser energies, a precise relationship is found between bias voltage $V_b$ and $X^0$ transition energy. This ability to precisely convert between $V_b$ and transition energy allows us to perform high-resolution measurements of the $X^0$ homogeneous linewidth and, thus, the electron tunnelling rate as a function of electric field, while taking into account power broadening of the $X^0$ linewidth for increasing resonant laser excitation intensity. In addition, we show that the measured electric-field-dependent electron tunnelling rate can be described accurately by a theoretical model based on a one-dimensional (1D) Wentzel-Kramers-Brillouin (WKB) approximation, yielding insight into the QD electron confinement potential and the QD height. Finally, having provided a comprehensive study of voltage-controlled electron tunnelling from a self-assembled QD to a 2DEG and shown that our device can operate as a 2DEG-based QD photovoltaic cell, we then provide two proposals for future optical spintronic devices – a QD spin-transistor and a QD photovoltaic spin-injector.

## II. EXPERIMENTAL DETAILS

As shown schematically in Fig. 1(a), the device used throughout this work was designed for PC measurements on single QDs and fabricated as an *n-i*-Schottky photodiode structure based on a 2DEG. The single layer of InAs self-assembled QDs, which was grown to yield a low surface-density of QDs (~$10^9$ cm$^{-2}$), is embedded in a 250-nm-thick *i*-GaAs layer and located 50 nm above a Si $\delta$-doped GaAs layer ($N_d$ = 5 x $10^{12}$ cm$^{-2}$), from which the 2DEG forms that is confined in the resultant V-shaped potential well.[8,9] Submicrometer-sized apertures were etched into an Al shadow mask via electron-beam lithography and serve to isolate single QDs for our single-QD PC measurements. Cr/Au bond pads were fabricated on the ohmic and Schottky contacts to allow for electrical connection to a voltage source and current meter. Further details on the design and fabrication of our device can be found in Ref. 2.

Figure 1(b) presents theoretical calculations of the band-edge diagram, in addition to the QD and 2DEG eigenstates, for our 2DEG-based *n-i*-Schottky photodiode using a 1D self-consistent Poisson-Schrödinger solver. According to the results for the conduction band, as shown in the inset of Fig. 1(b), it is indeed possible to tune the QD *s*-shell electron level above the quasi-Fermi level $E_F$ in our device using a suitable $V_b$. Specifically, for $V_b = 0$ V, the QD electron level is positioned energetically well above $E_F$ such that an optically-generated electron should tunnel quickly out of the QD prior to *e-h* recombination and thus contribute towards a PC through the external electronic circuit, as will be confirmed experimentally in the following section. Note that according to these band-edge calculations, the QD *s*-shell electron level is positioned above $E_F$ under zero-bias conditions ($V_b = 0$ V). Therefore, this suggests that our device may also operate as a 2DEG-based QD photovoltaic cell, as will be confirmed by the single-QD PC measurements in the following section. For this calculation, it was assumed that the QD *s*-shell exciton transition energy is ~1.34 eV, which is within the energy range of our QD ensemble, and that the QD height is 5 nm (Ref. 2).

Figure 1(c) shows theoretical calculations of the conduction band and eigenstates of our device's 2DEG, which is derived from Si $\delta$-doping. The results illustrate a V-shaped conduction-band profile, which is the result of a $\delta$-function-like distribution of positive ionized donors, with four confined two-dimensional (2D) quantized subbands whose eigenvalues $E_0$, $E_1$, $E_2$, and $E_3$ are shown together with their respective wave function probability distributions. It is important to note that, especially for the higher-order subbands, the spatial extent of the delocalization of electrons from their ionized donors and the spreading-out of electronic states is much larger (~20 nm) compared to that of the localized $\delta$-function-like ionized donor profile, which is similar to the GaAs lattice constant.[8] Therefore, electron mobilities and the effects of ionized impurity scattering in such 2DEGs formed from $\delta$-doping is expected to be similar to those in 2DEGs formed from modulation-doped heterostructures.[10]

In the single-QD PC measurements performed throughout this work, resonant laser excitation was derived from a tunable narrow-bandwidth (~1 MHz) external-cavity diode laser (ECDL) in the Littrow configuration. After passing it first through a combination of a half-wave plate, a polarizing beamsplitter, and a second half-wave plate to control its intensity and linear polarization, the output of the laser was focussed on a given aperture of a device using a microscope objective lens [numerical aperture (NA) = 0.5], which was mounted on a piezo-driven *xyz*-stage for precise nano-positioning. Experiments on the devices were performed at low temperatures (~4.2 K) in a He-flow cold-finger optical cryostat, while $V_b$ and measurement of the PC signal were provided by a commercial source-measure unit (SMU). Finally, a micro-photoluminescence ($\mu$-PL) spectroscopy system, as described in Ref. 2, was used for initial bias-dependent $\mu$-PL measurements. This was necessary in order to determine promptly the relevant $X^0$ transition energy and $V_b$ range for subsequent PC measurements on a given single QD, as will be shown in the next section.

### III. RESULTS AND DISCUSSION

Prior to carrying out the required single-QD PC measurements for this study, initial bias-dependent $\mu$-PL spectroscopy was performed on several single QDs, each isolated in an aperture with a diameter of ~300 nm, in order to determine quickly their particular $X^0$ transition energy and relevant $V_b$ range corresponding to the PC measurement regime. Fig. 2 presents a contour plot of the time-integrated $\mu$-PL spectra from one particular QD for the bias range of 0.35 V ≤ $V_b$ ≤ 0.85 V with a laser excitation intensity of ~4.5 $\mu$W. This plot is characterized by the appearance of four distinct PL emission lines over the measured $V_b$ range, each representing a different QD excitonic state with a unique transition energy due to the given Coulomb interactions between charge carriers confined in the QD. Through laser-intensity-dependent measurements of the integrated PL intensity for each of the PL emission lines observed in Fig. 2, it was confirmed that the emission lines at ~1.372, ~1.366, and ~1.360 eV originate from single-exciton-states, while the emission line at ~1.367 eV originates from the neutral two-exciton state occupying the QD $s$-shell, known as the biexciton state (2X). In our device, the vertical electric field $F$ at the QD layer is defined as[8,10] $F = (V_i - V_b)/d$, where $V_i$ is the intrinsic built-in potential (Schottky barrier) and $d$ is the distance between the Si $\delta$-doping and Schottky contact. $V_i$ was determined in our device by measuring the value of $V_b$ at which the PC signal changed sign (i.e., at the flat-band condition) for non-resonant laser excitation using a HeNe laser. It is also important to note that motivations for choosing $\delta$-doping for this study, as opposed to other approaches (eg., a modulation-doped heterostructure), to obtain the 2DEG in our device were the ease of determining the electric field at the QD layer and the linearity of the electric field as a function of bias voltage.[8,10]

Following our discussion in Ref. 2 on a similar QD, we are able to assign the PL emission lines at ~1.366 and ~1.360 eV to $X^0$ and the singly negatively-charged exciton ($X^-$), respectively, in the bias-dependent $\mu$-PL spectra of Fig. 2. Therefore, the contour plot is divided into three bias ranges: $V_{PC}$ ($V_b$ < 0.46 V), where no PL emission is observed and instead a PC is generated, $V_0$ (0.46 V ≤ $V_b$ < 0.68 V), where $X^0$ PL emission is observed, and $V_-$ ($V_b$ ≥ 0.68 V), where $X^-$ PL emission appears due to bias-controlled single-electron charging of the QD by the nearby 2DEG. The relatively weak PL emission line at ~1.372 eV is assigned to the singly positively-charged exciton ($X^+$). Here, the formation of $X^+$ occurs within the hole tunnelling time as an optically-generated *e-h* pair is added to a QD charged with a single hole as a result of its $X^0$ electron tunnelling out faster than the $X^0$ recombination time, given that the two holes have opposite spins. This interpretation is supported by the fact that the appearance of the 2X line at ~1.367 eV coincides with the decrease in $X^+$ PL intensity. It is important to note here that any spin-polarization of optically-excited carriers in the GaAs continuum of states is lost upon their capture and relaxation into the QDs.[2] As seen in Fig. 2, the presence of an additional positive charge carrier in the QD results in a blue-shift of the $X^+$ transition energy relative to that of $X^0$, due to the *h-h* Coulomb repulsion outweighing the additional *e-h* attraction in the $X^+$ few-particle state.[11] Compared to those reported in other works,[12-14] the large blue-shift of the $X^+$ transition energy for this QD (~6 meV) is attributed to a reduced effect of correlation, which is a result of a decrease in the number of bound hole-states in this small QD.[11]

It is worth noting in Fig. 2 that this particular QD exhibits a negative 2X binding energy, as opposed to the positive 2X binding energies observed in most other works on similar QDs.[2,12,15] This transition from "binding" to "antibinding" is due to a decrease in the number of localized excited states in this relatively small QD, as indicated from its comparatively high exciton transition energy, reducing the contributions of correlation and exchange towards the 2X binding energy.[16]

As was predicted in the band-edge calculations of Sec. II, the absence of PL emission at $V_b$ = 0 V reveals that the electron tunnelling rate is much faster than the exciton recombination rate when our photodiode is unbiased. Therefore, as will be confirmed in the PC measurements presented in the following subsection, it is possible that our device may function as a 2DEG-based QD photovoltaic cell.

So far in this section, we have examined bias-dependent $\mu$-PL spectroscopy on a single QD in order to determine its particular $X^0$ transition energy and the $V_b$ range within which a PC is expected to be generated following resonant laser excitation of $X^0$. As will be demonstrated throughout the remainder of this work, such device characterization will then allow us to perform the desired single-QD PC measurements under high-electric-field conditions.

### A. Single-QD PC spectrum and the QCSE

The single-QD PC spectrum of $X^0$ is obtained by sweeping the $X^0$ transition energy through a known fixed laser energy $E_{laser}$ via the QCSE, which is here a result of an electric-field-dependent exciton dipole moment in the QD along the growth direction. In such a measurement, when the $X^0$ absorption peak is tuned into resonance with $E_{laser}$ within the established high-electric-field $V_{PC}$ bias range, resonant excitation of an $X^0$ exciton in the initially empty QD is followed by exciton ionization through rapid electron and hole tunnelling out of the QD towards the 2DEG and Schottky contact, respectively, thereby contributing towards a PC signal that is measured as a DC current through an external electric circuit. A typical such single-QD PC spectrum of $X^0$ is presented in Fig. 3 for $E_{laser}$ = 1364.23 meV and an incident laser-excitation intensity on the sample of 400 nW. The $X^0$ exciton state in our QDs is characterized by two orthogonal linearly-polarized eigenstates aligned to the [110] crystal axes that is a result of an anisotropic $e$-$h$ exchange interaction.[17] As a result, the excitation laser field was chosen to be linearly-polarized and aligned parallel to the sample's [110] crystal axis for all measurements throughout this work. To perform this initial PC measurement, since the QCSE has the effect of red-shifting the exciton transition energies under high-electric-field conditions characteristic of the $V_{PC}$ bias range, this value of $E_{laser}$ was chosen arbitrarily to be slightly red-shifted relative to the $X^0$ transition energy observed in the previous $\mu$-PL measurements. Now, by fitting a Lorentzian curve (solid line in Fig. 3) to this PC peak and extracting its central bias voltage, we are thus able to determine the corresponding values for $V_b$ and $X^0$ transition energy for this particular QD and photodiode device.

It is important to point out that this method of sweeping the $X^0$ transition energy through a fixed $E_{laser}$ via the QCSE, as opposed to the opposite method of sweeping the energy of a tunable laser through a fixed $X^0$ transition energy for a given $V_b$, has the advantage of avoiding possible mode-hopping of our ECDL if $E_{laser}$ is tuned beyond its rated mode-hop-free tuning range (~20 GHz) during the measurement of a single-QD PC spectrum. Furthermore, the ease of sweeping $V_b$ to acquire a PC spectrum (Fig. 3) greatly reduces the experiment measurement time and simplifies the overall data acquisition process.

In order to be able to extract values for the QD homogeneous linewidth and hence the electron tunnelling rate from the single-QD PC spectrum in Fig. 3, it is necessary to first convert the measurement from a $V_b$ spectrum to an equivalent energy spectrum. This is achieved by performing the equivalent single-QD PC measurement in Fig. 3 for a sequence of different known values of $E_{laser}$, as the $X^0$ absorption peak will appear at different values of $V_b$ for varying $E_{laser}$ according to the QCSE. Then, using the resultant corresponding values of $V_b$ and $X^0$ transition energy, which is known from the corresponding values of $E_{laser}$, it will be possible to interpolate precise values for $X^0$ transition energy as a function of $V_b$. Accordingly, Fig. 4 presents single-QD PC spectra for a series of distinct values of $E_{laser}$, which are indicated alongside their corresponding spectrum, throughout the bias range within which a measurable single-QD PC signal was obtained. The spectra are shifted vertically with respect to each other for clarity and the laser excitation intensity used here was 2 $\mu$W. Each PC spectrum is fit with a Lorentzian curve (solid lines in Fig. 4) to determine its peak $V_b$. While moving from high to low values of $V_b$, we first observe (Fig. 4) an increase in the PC peak amplitude, which is due to an increasing hole-tunnelling efficiency for increasing electric-field conditions, followed by a gradual decrease in the PC peak amplitude until eventually a peak is no longer visible. This decrease in peak amplitude towards high-electric-field conditions is the result of broadening of the $X^0$ absorption peak as the exciton lifetime decreases, in accordance with time-energy uncertainty, due to a rapidly increasing electron tunnelling rate out of the QD. It should also be noted in Fig. 4 that the PC spectra towards high values of $V_b$ demonstrate a measurable PC signal at $V_b = 0$ V, thereby confirming the ability of our device to operate as a 2DEG-based QD photovoltaic cell.

Figure 5 shows a plot of $X^0$ transition energy as a function of $V_b$ obtained through both PC and PL spectroscopy of our single QD. While the data points in the PC regime were acquired by extracting values of the peak $V_b$ from Lorentzian curve fits to single-QD PC spectra (Fig. 4), those in the PL regime were acquired by extracting values of the peak energy from Lorentzian curve fits to the single-QD $\mu$-PL spectra in Fig. 2. Note the relatively large difference between the value of $V_b$ at which quenching of $X^0$ PL emission is observed and the appearance of a measurable $X^0$ PC peak, compared to other works.[18,19] We attribute this to a very long hole-tunnelling time in our QD such that a measurable single-QD PC spectrum is obtained only towards relatively high electric-field conditions, as suggested by the comparatively low PC peak amplitudes observed in Fig. 4 and the appearance of an $X^+$ emission line in the bias-dependent $\mu$

-PL spectra of Fig. 2. Such long hole-tunnelling times in our QD will be confirmed in the next section through an analysis of the saturation of the peak PC amplitude towards high laser-excitation intensities. From the QCSE, a quadratic curve can be fit to the data points in Fig. 5, yielding a precise relation that is used to convert between $V_b$ and $X^0$ transition energy in this particular QD and device (Fig. 5).

### B. Power broadening of $X^0$ homogeneous linewidth

Having obtained the ability to precisely convert from $V_b$ to $X^0$ transition energy in the last subsection, the $X^0$ homogeneous linewidth $\Gamma_0$ and electron tunnelling rate can now be determined from a single-QD PC spectrum (Fig. 3). However, in order to derive $\Gamma_0$, we must first take into account power broadening,[20] which produces a broadening of the $X^0$ absorption peak for increasing resonant laser excitation intensity. A commonly-known phenomenon in atomic optics, power broadening of $\Gamma_0$ is the result of a saturation effect of the single-QD PC signal towards high laser-excitation intensities that is determined by the slow hole-tunnelling time, since the $X^0$ transition cannot be excited until all carriers have tunnelled out of the QD as a result of a renormalization of exciton transition energies due to Coulomb interactions in the exciton few-particle states. Therefore, while the magnitude of the PC signal at exact resonance to the $X^0$ absorption peak is close to its saturation value and approaching this value asymptotically, the magnitude of the PC signal increasingly detuned from resonance (i.e., towards the edges of the PC peak) is well below its saturation and thus will experience a higher rate of increase compared to that at exact resonance, resulting in a broadening of the $X^0$ absorption peak.

The saturation of the PC amplitude for high laser-excitation intensities is demonstrated in Fig. 6, where the peak PC amplitude $I_{peak}$ is plotted as a function of laser excitation intensity for an $X^0$ absorption peak centered on $V_b$ = -0.03 V ($F$ = 37.2 kV/cm). This saturation effect in our QD $X^0$ two-level system can be described by the following theoretical model[21]

$$I_{peak} = \frac{e}{2\tau_T^h} \frac{\tilde{P}}{\tilde{P}+1} = I_{sat} \frac{\tilde{P}}{\tilde{P}+1} \tag{1}$$

where $I_{sat}$ is the saturation peak PC amplitude for a given $V_b$ or $F$, $\tau_T^h$ is the hole tunnelling time, $e$ is the elementary charge, and $\tilde{P}$ is the normalized laser-excitation intensity on the sample. As seen by the solid line in Fig. 6, fitting this model to the experimental data yields $I_{sat} = 14.2 \pm 0.1$ pA and thus $\tau_T^h = 5.64 \pm 0.05$ ns, given that the laser excitation intensity was normalized to 1 $\mu$W on the plot's *x*-axis in order to perform an accurate fit to the experimental data.

Having shown that $I_{peak}$ approaches a saturation value towards high laser intensities, the magnitude of a PC signal on the sides of the absorption peak and increasingly detuned from exact resonance should increase at a higher rate than that at resonance for increasing laser intensity, thereby resulting in a broadening of the $X^0$ absorption peak. Such power broadening can be seen from the series of PC spectra in the inset of Fig. 6, from which were obtained the corresponding experimental data for $I_{peak}$ shown in the main part of the figure, for a range of laser excitation intensities that are indicated to the right-hand side of each spectrum. Also, while the spectra are shifted vertically with respect to each other for clarity, the top *x*-axis shows the equivalent values for $X^0$ transition energy, which were determined using the experimentally-derived equation in Fig. 5 relating $V_b$ and $X^0$ transition energy. It is worth noting the slight shift of the PC peaks towards lower $V_b$ for increasing laser excitation intensity that is attributed to electrostatic shielding effects,[22] as the number of optically-excited electrons and holes tunnelling to the 2DEG and Schottky contact, respectively, increases with increasing laser intensity. Consequently, after fitting the experimental data in the inset of Fig. 6 using Lorentzian curves (solid lines), the value of $V_b$ corresponding to the PC resonance peak observed in these spectra (i.e., $V_b$ = -0.03 V) was determined from the spectrum obtained using the lowest laser-excitation intensity.

The power-broadened linewidth $\Gamma$ of the $X^0$ absorption peak shown in the inset of Fig. 6 as a function of laser excitation intensity (Fig. 7) can be described by[20,22]

$$\Gamma = \Gamma_0 \sqrt{1+\tilde{P}}. \qquad (2)$$

As shown by the solid line in Fig. 7, the above theoretical model fits well to the experimental data with the same normalization factor used in Fig. 6. Therefore, since the experimental data and theoretical models of Eq. (1) and Eq. (2) show close agreement in Fig. 6 and Fig. 7, respectively, we conclude that the effect of PC saturation is the *only* cause of the observed laser-power-dependent broadening of $\Gamma_0$ and, therefore, that the $X^0$ dephasing time $T_2$, which can be derived from $\Gamma_0$ since $T_2$ is limited by the fast electron tunnelling time, is independent of the laser excitation intensity.[22] By extrapolating the fit in Fig. 7 to a laser intensity of zero, we find that $\Gamma_0 = 156.7 \pm 0.7$ $\mu$eV for $V_b$ = -0.03 V ($F$ = 37.2 kV/cm) in this particular QD and device.

### C. Electron tunnelling time as a function of *F*

We now repeat the measurements presented in the previous subsection to obtain $\Gamma_0$ for a series of values of $V_b$ throughout the bias range within which a PC peak can be measured and fit using a Lorentzian curve, thereby yielding experimental data for $\Gamma_0$ as a function of *F*. The results of such a series of measurements is presented in Fig. 8, showing $\Gamma_0$ (squares) and the corresponding electron tunneling time $\tau_T^e$ (circles) as a function of $V_b$ and its corresponding $F$. Since the lifetime of $X^0$ is limited by fast electron tunnelling out of the QD, $\tau_T^e$ was derived from $\Gamma_0$ via $\tau_T^e = \hbar/2\Gamma_0$. The electron tunnelling rate $R_T^e$ out of the QD can be modeled via a 1D (along the growth direction) WKB approximation[23]

$$R_T^e = \frac{\hbar \pi}{2m_e^* H^2} \exp\left[\frac{-4}{3\hbar eF}\sqrt{2m_e^* E_b^3}\right] \quad (3)$$

where $H$ is the QD height, $m_e^* = 0.067 m_e$ is the electron effective mass in GaAs (Ref. 24) and $m_e$ is the electron mass in vacuum, and $E_b$ is the tunnel barrier height. As can be seen from the solid lines in Fig. 8, the above theoretical model fits remarkably well to the experimental data throughout the range of $F$ within which a PC peak can be measured. Therefore, we are able to deduce precise values for $H$ and $E_b$ as a result of the model fitting (Fig. 8). Since the measurements throughout this work were performed at low temperatures (~4.2 K), carrier escape from the QD is the result of direct tunnelling from a bound QD s-shell state to the GaAs conduction or valence band through the triangular tunnel barrier, rather than the result of thermionic emission or thermally-assisted tunnelling via the InAs wetting layer (WL) states,[25] which are higher in energy relative to those of the QD s-shell due to the stronger quantum confinement along the growth direction in this 2D WL. Hence, we define $E_b$ in Eq. (3) to be the height of the GaAs triangular tunnel barrier for electrons tunnelling directly out of the QD.

The inset of Fig. 8 shows theoretical calculations of the conduction band and QD s-shell electron eigenstate using a 1D self-consistent Poisson-Schrödinger solver. Using $H = 4.4$ nm and an $X^0$ transition energy of 1.364 meV, the results of the calculation yield a tunnel barrier height of 58.9 meV for the QD s-shell electron. This value agrees reasonably well with the value of $E_b$ derived from the fit shown in the main part of Fig. 8 ($E_b = 52.2 \pm 0.3$ meV), supporting the accuracy of the theoretical model for $R_T^e$ in Eq. (3).

Figure 9 shows $\log(R_T^e)$ as a function of $F^{-1}$ obtained from the same experimental data and model fit as presented in Fig. 8. Here, however, the theoretical model fit using Eq. (3) has also been extrapolated to determine the $V_b$-value at which $R_T^e$ intersects the electric-field-independent exciton recombination rate $R_{PL} = 1$ ns$^{-1}$ (Ref. 26). Shown in the figure to be $V_b = 0.46$ V ($F = 17.6$ kV/cm), this is the threshold value of $V_b$ at which quenching of $X^0$ PL emission is expected according to the theoretical model, since there is a higher probability of exciton recombination (electron tunnelling) for $V_b$ above (below) this value. Now, comparing Fig. 9 with Fig. 2, we see that this $V_b$-value is in close agreement with that at which quenching of $X^0$ PL emission is observed in the bias-dependent $\mu$-PL spectra of Fig. 2, further supporting the experimental results and accuracy of the theoretical model of Eq. (3).

### D. Proposals for spintronic devices

In this article, we have provided a comprehensive investigation of voltage-controlled electron tunnelling from a single self-assembled QD to a 2DEG in a photodiode device structure, which we showed can also function as a 2DEG-based QD photovoltaic cell. Using a device structure similar to that presented in this work, we now propose two device concepts for applications in spintronics that may be made feasible as a result of this investigation. It is important to note that the proposed devices that are coupled to optical fields are made possible only through the use of self-assembled QDs, which are known to exhibit a high optical quality due to their relatively defect-free crystal structure.

(a) *QD spin-transistor (QDST)*. As seen in Fig. 10(a), the structure of the QDST is similar to that of the spin-SET that we proposed in Ref. 2, except for two important improvements that are made possible through the results obtained in the present work. First, while the spin single-electron transistor (spin-SET) required a ferromagnetic (FM) top gate to Zeeman-split the QD electron spin-states and thus increase the spin relaxation time to a value much longer than $\tau_T^e$ such that the electron would not undergo a spin-flip while resident in the QD, the experimental and supporting theoretical results reported in this work demonstrate that, for a suitable design of $H$ and $E_b$, $\tau_T^e$ can be made more than two orders of magnitude shorter than the electron spin relaxation times reported in similar QDs under zero magnetic fields,[27] thereby eliminating the need for an FM top gate. Therefore, as shown in Fig. 10(a), this permits us to use a *p*-doped semiconductor top gate in the QDST. Second, our choice of using a *p*-doped semiconductor top gate thus allows us to design the QDST based on two lateral *p-i-n* junctions, which are located at the etched edges of the mesa in Fig. 10(a) since the 2DEG is depleted for areas above which the *p*-doped layer exists.[3,28,29] Note that, while the spin-SET described in Ref. 2 requires a suitably large negative gate voltage $V_g$ to deplete the 2DEG below the QD such that the conductivity in the 2DEG channel below the QD is zero, the depleted area of the 2DEG under the QD in this QDST is determined simply by the presence of the *p*-doped top gate. Therefore, unlike the case of the spin-SET, designing the QD energy levels in the QDST is not crucial in terms of ensuring that both depletion of the 2DEG and QD-2DEG tunnel-coupling are maintained while the *s*-shell electron state is tuned through the Fermi levels of the source ($E_{FS}$) and drain ($E_{FD}$) as $V_g$ is modulated. Further, compared to using a metallic top gate, using a semiconductor top gate renders the optical-addressing of QDs in the vertical direction more practical. Energy-level diagrams of the QDST in the *x*-direction are shown in Fig. 10(b). As shown in the top diagram of Fig. 10(b), for an FM source (drain) that injects (transmits) spin-up electrons and a positive drain-to-source voltage $V_{ds}$, the QDST is turned on (i.e., non-zero drain-to-source current $I_{ds}$) if the QD *s*-shell electron state is energetically-positioned below $E_{FS}$ and above $E_{FD}$. Here, a spin-up electron is thus permitted to tunnel from source to drain via the QD, while no spin relaxation occurs in the QD given that the electron tunnelling rate is designed to be much faster than the spin relaxation rate in the QD. The QDST can then be gated-off electrically by tuning $V_g$ such that the QD electron state is positioned well above $E_{FS}$, resulting in zero $I_{ds}$ since no electrons are permitted to tunnel into the QD and hence to the drain. Alternatively, the QDST can be gated-off *optically* by inducing a $\pi$-rotation on the QD electron spin using an intense ultrafast laser pulse via the optical Stark effect,[30] leading to zero $I_{ds}$ as spin-down electrons would reflect, rather than transmit, upon arrival at the FM drain. Finally, although Fig. 10 describes the operation of a QDST for spin-up electrons, it is worth noting that the same operation can be demonstrated for spin-down electrons.

(b) *QD photovoltaic spin-injector*. As seen in Fig. 11(a), the structure of the QD photovoltaic spin-injector is similar to that of the device for optical spin-injection and -detection that we proposed in Ref. 2, except for several improvements that are made possible by the results reported in this Article. First, having demonstrated in this work that our 2DEG-based single-QD photodiode is capable of generating a PC at zero-bias conditions (i.e., $V_b = 0$ V), the device proposed here can thus operate as a photovoltaic cell to be used for optical spin injection within a spintronic circuit. Second, similar to the case of the QDST, the FM top gate is replaced by a *p*-doped semiconductor layer, since it has been demonstrated in this work that $\tau_T^e$ can be made more than two orders of magnitude shorter than the exciton spin relaxation times under zero magnetic fields[30] such that the exciton would not undergo a spin-flip before the electron tunnels out of the QD. Third, as a result of using a *p*-doped layer, a lateral *p-i-n* junction forms at the etched edge of the mesa in Fig. 11(a),[3,28,29] thereby offering the same advantages as those discussed above for the case of the QDST. The energy-level diagram of the QD photovoltaic spin-injector in the *x*-direction is shown in Fig. 11(b). To perform optical spin-injection using this device, the QD electron (hole) state is energetically-positioned above (below) the Fermi level in the 2DEG (*p*-doped layer) for zero-bias conditions between the 2DEG and *p*-doped layer. Then, for injection of a spin-up electron, an exciton is resonantly-excited in the QD using $\sigma^-$ circularly-polarized light to create a spin-up (-down) electron (hole). This can be achieved, for example, by using an ultrafast laser $\pi$-pulse that performs an inversion of the QD exciton two-level system.[32,33] Subsequently, the spin-up electron is injected into the 2DEG as it tunnels out of the QD, while the hole tunnels out towards the *p*-doped layer. Although Fig. 11 shows the operation of a QD photovoltaic spin-injector for spin-up electrons, the same can be achieved for optical injection of spin-down electrons. Of course, while we have proposed operating it as a photovoltaic cell, it is worth noting that this optical spin-injector design may be used also under biased conditions.

Finally, it is important to note that in the spintronic devices proposed in this section we have assumed ideal ferromagnetic contacts, which inject or transmit only the majority spins, and the absence of scattering-induced spin relaxation in the 2DEG due to spin-orbit coupling.

## IV. CONCLUSIONS

We have demonstrated voltage-controlled electron tunnelling from a single self-assembled QD embedded in the intrinsic region of a 2DEG-based $n$-$i$-Schottky photovoltaic cell. A high-resolution PC spectrum of $X^0$ was obtained by sweeping the $X^0$ transition energy through that of a fixed narrow-bandwidth laser via the QCSE. By repeating this measurement for a sequence of $E_{laser}$, a precise conversion between $V_b$ and $X^0$ transition energy was thus obtained. This ability then allowed us to determine $\Gamma_0$ and, hence, $\tau_T^e$ as a function of $F$, while taking into account power broadening of the $X^0$ absorption peak, which was confirmed to be due to a saturation effect of the PC from a single QD. We found that the $F$-dependent $\tau_T^e$ can be described accurately by a theoretical model based on a 1D WKB approximation, yielding insight into the QD electron confinement potential and the QD height. Finally, having offered a detailed study of voltage-controlled electron tunnelling from a single self-assembled QD to a 2DEG and shown that our device can also function as a 2DEG-based QD photovoltaic cell, we proposed two spintronic device concepts that may be made feasible as a result of our investigation.

## ACKNOWLEDGEMENTS

J. D. M. gratefully acknowledges financial support from HEFCE, CCT, and NSERC (PGS M, PGS D).

**FIGURES**

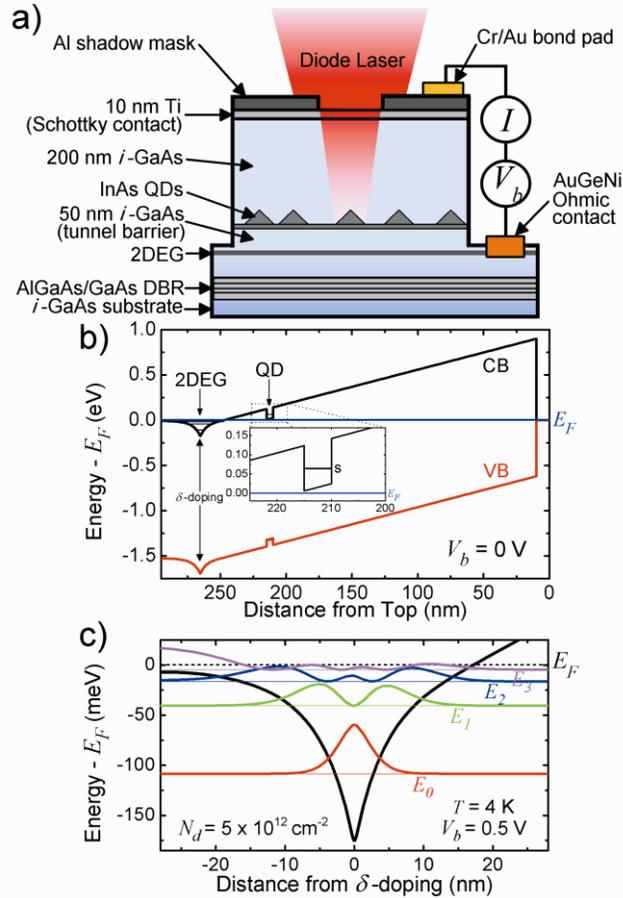

FIG. 1. (Color online) (a) Schematic diagram of the 2DEG-based $n$-$i$-Schottky photodiode used for voltage-controlled electron tunnelling from single QDs in single-QD PC spectroscopy. (b) Theoretical calculations of the conduction band, valence band, and QD and 2DEG eigenstates for the $n$-$i$-Schottky photodiode, showing the ability to tune the QD $s$-shell electron state above the Fermi energy $E_F$ at zero-bias conditions ($V_b = 0$ V), as seen in the inset. (c) Theoretical calculations of the V-shaped conduction band profile and 2DEG eigenvalues ($E_0$, $E_1$, $E_2$, $E_3$) and the respective probability distributions for the four confined 2D subbands, which are the result of the Si $\delta$-doping ($N_d = 5 \times 10^{12}$ cm$^{-2}$).

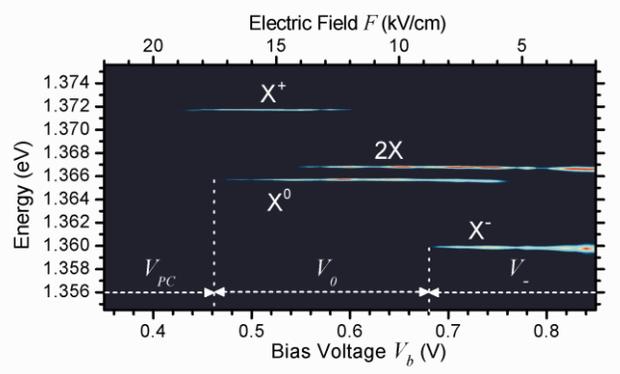

FIG. 2. (Color online) Contour plot of the bias-dependent $\mu$-PL spectra of a single QD, showing PL emission lines from $X^+$, $X^0$, $X^-$, and 2X.

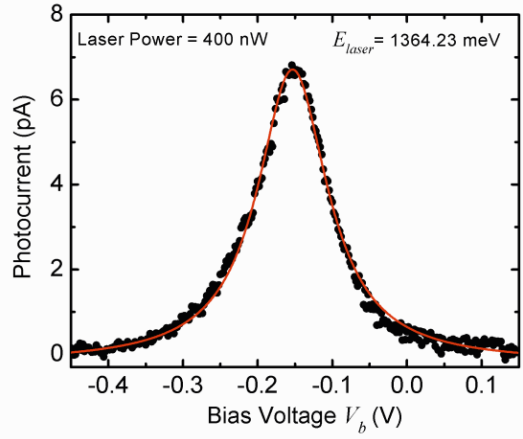

FIG. 3. (Color online) Single-QD PC spectrum of $X^0$ for $E_{laser}$ = 1364.23 meV and a laser excitation intensity of 400 nW. The solid line is a Lorentzian fit curve to the experimental data, yielding a peak $V_b$ of -153.2 ± 0.3 mV.

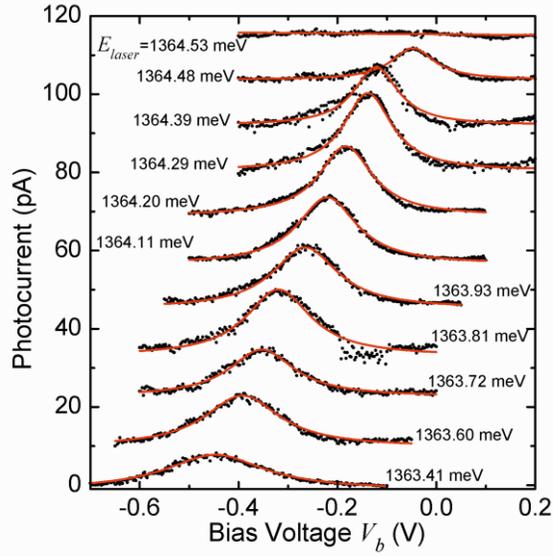

FIG. 4. (Color online) Single-QD PC spectra of $X^0$ for a series of values of $E_{laser}$ with Lorentzian fit curves (solid lines). For clarity, spectra are shifted vertically with respect to each other.

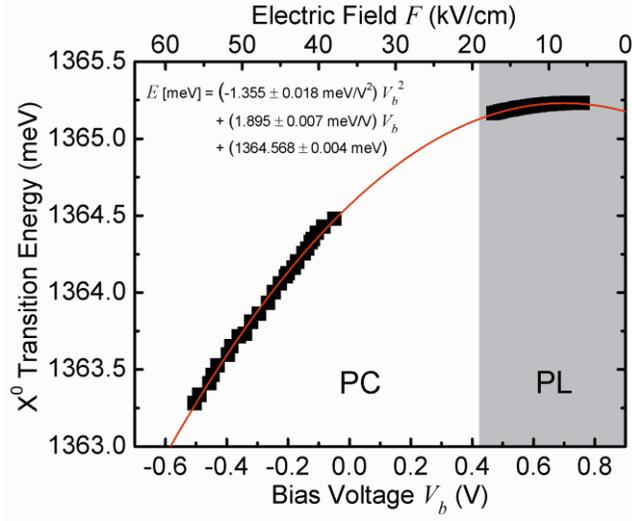

FIG. 5. (Color online) Plot of $X^0$ transition energy as a function of $V_b$ and $F$, obtained from Lorentzian fits to the single-QD PC and $\mu$-PL spectra. As a result of the QCSE, a quadratic curve (solid line) is fit to the experimental data for $X^0$ transition energy vs. $V_b$. The result of the quadratic fit is shown in the figure.

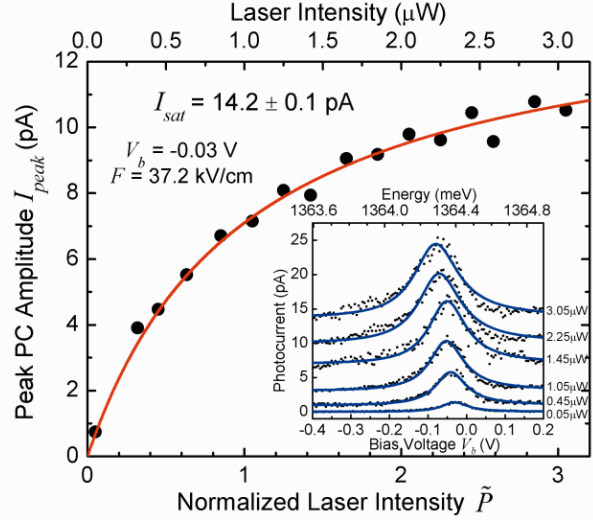

FIG. 6. (Color online) Plot of the peak PC amplitude $I_{peak}$ as a function of normalized laser-excitation intensity $\tilde{P}$ for an $X^0$ absorption peak centered on $V_b$ = -0.03 V ($F$ = 37.2 kV/cm). The solid line is the fit curve for $I_{peak}$ described by Eq. (1), yielding a saturation peak PC amplitude $I_{sat} = 14.2 \pm 0.1$ pA and a hole tunnelling rate $\tau_T^h = 5.64 \pm 0.05$ ns. Inset: Single-QD PC spectra for a series of laser excitation intensities, as indicated on the right-hand side of each spectrum, that were used towards the $I_{peak}$ data points shown in the main part of the figure. The top x-axis corresponds to the $X^0$ transition energy, as derived from the quadratic fit curve shown in Fig. 5. The spectra are shifted vertically with respect to each other and the solid lines are Lorentzian fit curves to the experimental data.

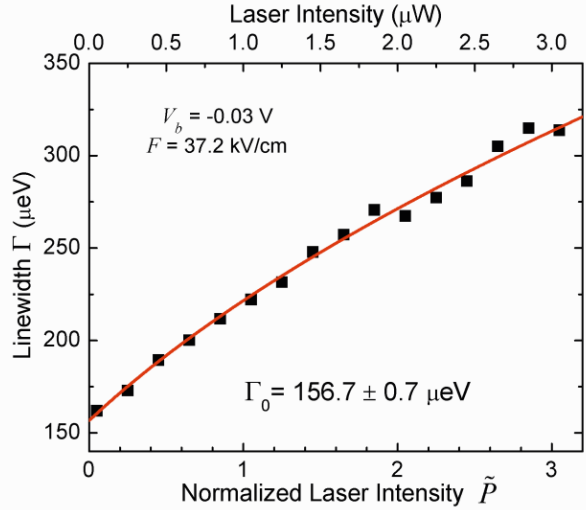

FIG. 7. (Color online) Plot of the power-broadened linewidth $\Gamma$ as a function of normalized laser-excitation intensity $\tilde{P}$ for an $X^0$ absorption peak centered on $V_b$ = -0.03 V ($F$ = 37.2 kV/cm). The solid line is the fit curve for $\Gamma$ described by Eq. (2), yielding the $X^0$ homogeneous linewidth $\Gamma_0 = 156.7 \pm 0.7 \mu$eV. A normalization factor of 1 $\mu$W was used here, consistent with that which was used to fit Eq. (1) to $I_{peak}$ in Fig. 6.

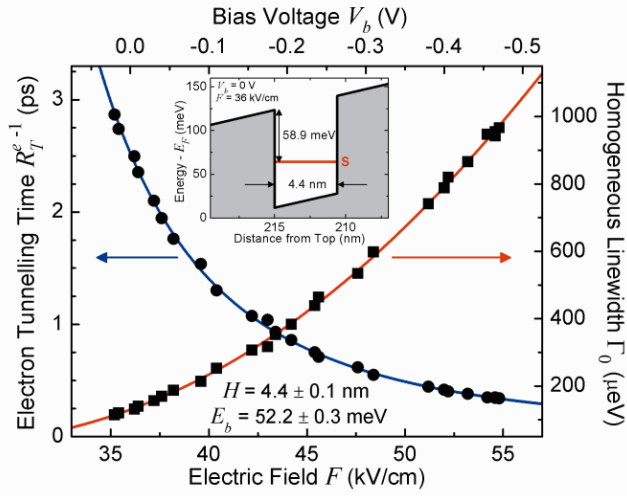

FIG. 8. (Color online) Plot of $\Gamma_0$ (squares) and electron tunnelling time $R_T^{e-1}$ (circles) as a function of $F$. The solid lines are fit curves using a theoretical model based on a 1D WKB approximation, which is defined in Eq. (3), yielding $H = 4.4 \pm 0.1$ nm and $E_b = 52.2 \pm 0.3$ meV. Inset: Theoretical calculation of the conduction band and QD *s*-shell electron eigenvalue using a 1D self-consistent Poisson-Schrödinger solver.

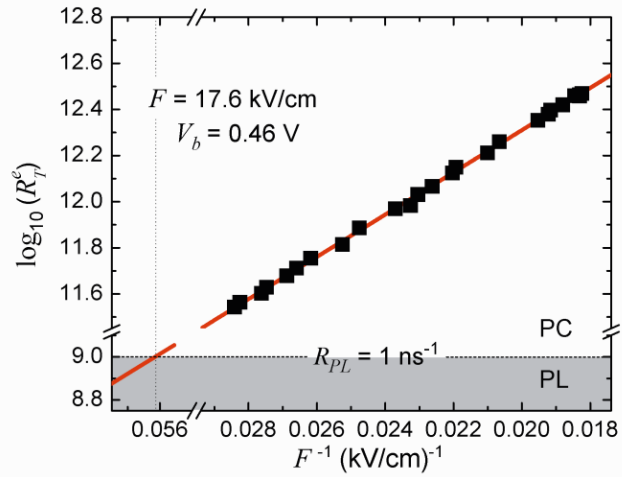

FIG. 9. (Color online) Plot of $\log\left(R_T^e\right)$ as a function of $F^{-1}$, where the theoretical fit curve (solid line) is extrapolated to show the point of intersection with $R_{PL}$ at $V_b = 0.46$ V. This value of $V_b$ is in agreement with that at which quenching of $X^0$ PL emission is observed in Fig. 2.

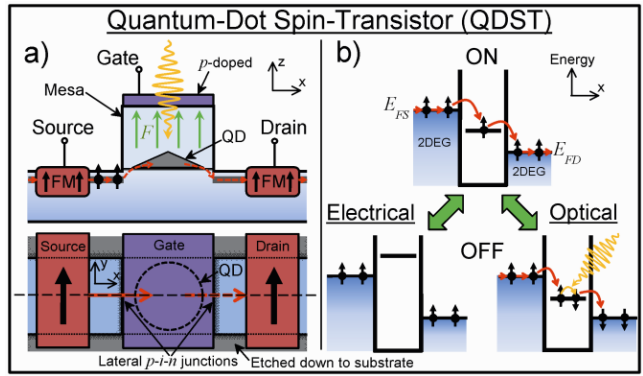

FIG. 10. (Color online) QD spin-transistor (QDST) proposal, showing a) the device structure and b) the corresponding energy-level diagrams for both electrical and optical gating. The red arrows indicate the presence of a spin current in the device.

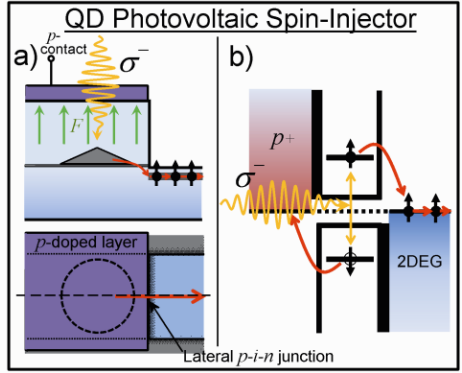

FIG. 11. (Color online) QD photovoltaic spin-injector proposal, showing a) the device structure and b) the corresponding energy-level diagrams for injection of spin-up electrons. Note that the device is photovoltaic since the photodiode creates a PC under zero-bias conditions. $\sigma^-$ indicates right circularly-polarized laser excitation.